\documentclass[aps,pra,showpacs,showkeys,onecolumn,superscriptaddress, notitlepage]{revtex4-1}

\usepackage{graphicx,tabularx,times,hyperref}
\usepackage{caption}
\setcitestyle{numbers,square}
\usepackage{float}

\usepackage{titlesec}

\titleformat*{\subsubsection}{\small\bfseries}

\usepackage{array}
\usepackage{amsmath}
\usepackage{units}
\usepackage{graphicx}
\usepackage{braket}
\usepackage{xfrac}
\usepackage{enumitem}
\usepackage[normalem]{ulem}
\usepackage{booktabs}
\usepackage{tabularx}
\usepackage{multirow, bigstrut}
\usepackage{hhline}
\usepackage{times}

\begin{document}

\title{Physics teaching assistants' views of different types of introductory problems: \\  Challenge of perceiving the instructional benefits of context-rich and multiple-choice problems}
\author{Melanie Good}
\author{Emily Marshman}
\affiliation{Department of Physics and Astronomy, University of Pittsburgh, Pittsburgh, PA 15260}
\author{Edit Yerushalmi}
\affiliation{Department of Science Teaching, Weizmann Institute of Science, 234 Herzl St., Rehovot, Israel 7610001}
\author{Chandralekha Singh}
\affiliation{Department of Physics and Astronomy, University of Pittsburgh, Pittsburgh, PA 15260}

\begin{abstract}
Physics problems can be posed in different ways.  Given a physics scenario, different problem types presenting that scenario in various ways can emphasize different instructional goals.
In this investigation, we examined the views of physics graduate teaching assistants (TAs) enrolled in a semester-long TA professional development course about the instructional benefits of different types of introductory problems based upon the same problem scenario to generate discussion and reflection on their use in different instructional situations. The TAs were asked to list the pros and cons of the problem types, rank them in terms of their instructional benefit and the level of challenge they might produce for their students, and describe when and how often they would use different types of problems in their own classes if they had complete control of teaching the class. Here we report on TAs' views about two of these problem types that were regarded by TAs as the least instructionally beneficial of all problem types--the context rich  and  multiple-choice formats. Many TAs listed no pros at all for these problem types, despite being explicitly asked for at least one pro. 
  They viewed multiple-choice questions nearly exclusively as tools for high-stakes summative assessment rather than their possible use as formative assessment tools, e.g., as clicker questions even in large classes. Similarly, TAs viewed context-rich problems as overly challenging, unnecessarily wordy, and too time-consuming to be instructionally beneficial to their students.  
 It is possible that in the written responses, TAs could have focused on the example problems provided to illustrate each problem type. Therefore, discussion in the TA professional development class and in the follow-up interviews explicitly included a focus on the general instructional benefits of well-designed multiple-choice and context-rich problems in different instructional contexts based upon the goals. It appears that TAs' sentiments were general views about these types of problems, and not just their views about the specific examples that the TAs were given in order to illustrate a problem type.
While TAs' concerns have obvious validity and value, 
the benefits of well-designed multiple-choice questions as a formative assessment tool was not readily identified by them, nor did the TAs recognize the learning benefits associated with solving context-rich problems.  Given the powerful ways multiple-choice and context-rich problems can be used for active engagement and formative assessment in different instructional contexts to meet diverse instructional goals, the lack of enthusiasm for these types of problems 
has implications for future TA professional development programs.

\end{abstract}

\maketitle

\vspace{-1.2cm}

\section{Introduction}
\vspace*{-.2cm}
\subsection{The development of expertise in introductory physics}
\vspace*{-.2cm}

The desired learning goals for students in many introductory physics courses often include learning  physics concepts and developing expertise in problem-solving and reasoning skills \cite{Yerushalmi_beliefs,prob0,prob1,prob2,prob3}. Physics experts, e.g., physics faculty members, organize their physics knowledge hierarchically so that underlying concepts are connected in a meaningful and structured way and they exhibit positive attitudes towards scientific problem solving \cite{Chi, Reif, Reif1,reif2,partial,Bolton,edit}. Experts' knowledge, including how the knowledge is structured in well-organized schema, and their positive attitudes to problem solving can facilitate an effective approach towards problem solving \cite{reif2,Singhc,Singhc1,Singhc2,Singhc3,Singhc4,Singhc5,Singhc6,Singhc7,Singhc7a}.  By contrast, many introductory students view physics as a collection of disconnected facts and equations and they have less expertlike attitudes towards problem-solving \cite{Chi, Reif, reif2}.  One strategy to achieve the goals related to the development of expertise of introductory physics students and improving their attitudes to problem-solving is to actively-engage them in the learning process using research-based approaches.

Different problem ``types'' (i.e., different ways of posing the same underlying physics problem) can be used in different ways to actively engage students using research-based approaches in order to meet the instructional goals and help students develop expertise.   
Depending on the instructional goals, active engagement methods can include a wide variety of options to meet those goals \cite{Henderson,Hake}, and different types of problems can be utilized to support the goals.  For example, even in large-enrollment classes, multiple-choice questions can be used, e.g., they can be administered via clickers to help provide formative assessment opportunities and engage students in discussion with peers to improve learning.  Such formative assessment opportunities afforded by multiple-choice questions can also help students take ownership of their learning \cite{Nicol, Faust}.  Another example is the use of context-rich problems (problems posed in a realistic, narrative manner which may include extraneous information and may not explicitly ask a question, i.e., an explicit problem may need to be formulated by the problem-solver). A problem posed in a context-rich manner can engage students in learning effective problem-solving strategies when used as part of collaborative group problem-solving while the instructor or teaching assistant facilitates the process by providing feedback and support as needed.  Moreover, group problem-solving with context-rich problems and ``Think-Pair-Share'' activities using multiple-choice clicker questions can promote both positive inter-dependence among students as well as individual accountability \cite{Faust}. While many other problem types exist, we focus here on multiple-choice and context-rich problem types, their role in helping students learn physics, and the way these types of problems are perceived by physics graduate teaching assistants (TAs) enrolled in a TA professional development course. 

\vspace*{-.2cm}
\subsection{The role of teaching assistants in promoting and supporting student learning and theoretical framework for our research}
\vspace*{-.2cm}

Physics graduate teaching assistants are often employed, especially at large research universities, to carry out duties such as instructing the recitation/discussion sections related to introductory physics courses.  Moreover, TA professional development programs may be the only opportunity for growth as an instructor that TAs may have before becoming a future faculty.  Furthermore, it has been noted that even though TAs are often responsible for a significant portion of undergraduate instruction, their training for this role is often limited \cite{conf2008, Nyquist}.  Even though a small minority of physics departments in the U.S. provide semester-long TA professional development, the majority of physics departments provide only very short training (i.e., a few hours) to prepare them for these various teaching responsibilities. 

Some TAs may be responsible for choosing the types of problems to use with introductory students, e.g., in designing quizzes for their students to take during recitation/discussion or creating example problems to discuss.  Moreover, as potential future faculty, TAs may have an ongoing decision-making responsibility about the types of problems to use with their future students \cite{Goertzen, Chini, Maries,Maries1, Maries2,Maries3,Lawrenz,Singh3,Singh31,Singh32,Singh33,Singh34,Singh36}.   
Thus, with limited opportunities for professional development and training in the intervening time between the TA role and the faculty role, TAs' perceptions may also shed light on their perceptions as future faculty members. 

The theoretical framework that inspired this research is that it is important to first investigate TAs' perceived ideas about various aspects of teaching in order to provide them suitable professional development opportunities to reflect critically upon the purpose of instruction and the importance, e.g., of formative assessment and collaborative group problem solving in bridging the gap between teaching and learning. In particular, research shows that formative assessment, e.g., clicker questions, can provide timely feedback and play a central role in aligning the learning goals, instructional design and assessment even in large classes \cite{Mazur}. Similarly, working with peers, e.g., using context rich problems, can help students take advantage of their peers to learn \cite{Heller}.
However, since TAs are not blank slates, the manner in which they themselves have been instructed in the past can strongly influence their views about different aspects of teaching and learning. Since many physics instructors teach primarily using traditional lectures and a significant portion of the recitation time is spent with the TAs solving example problems on the board for introductory physics students, such experiences in their own undergraduate years can greatly shape TAs' views about how students should be taught in different instructional settings. 
For example, based upon their prior experiences,TAs may have perceived instructional value for different types of problems. These perceived values may affect choices about the use of various types of problems in different instructional settings to meet different instructional goals and thus impact the degree to which different types of problems will be exploited for their effectiveness in actively engaging students and their ability to facilitate different instructional goals. Therefore, TAs' perceptions of different types of problems are worthy of examination to inform professional development courses and programs to improve student learning. 

\vspace*{-.35cm}
\subsection{Focus of our research}
\vspace*{-.35cm}

In the study presented here, TAs in a TA professional development course  were asked to reflect upon different problem types and their features that may be appropriate for different instructional settings to meet different instructional goals in introductory physics.  These same problem types had been used in an earlier study with physics instructors \cite{Edit}. The example problem types given to the TAs were for illustration purposes and were meant to generate a broader general discussion and reflection upon those types of problems and whether the TAs discern instructional benefit and would want to use a particular problem type in different instructional settings to meet diverse instructional goals if they had full control of teaching the class. Here we focus on TAs' initial views, after some experience in their first semester as a TA, about the multiple-choice and context-rich introductory physics problem types and
investigate the following research questions:
(1) How challenging and instructionally beneficial do TAs perceive context-rich and multiple-choice problem types? 
(2) If TAs had complete control of teaching an introductory course, how likely would they be to use these types of problems compared with other types of problems, and for what purpose might they use them?
(3) Why do TAs perceive context-rich and multiple-choice problem types the way they do? 

\vspace*{-.35cm}
\section{Background}
\vspace*{-.3cm}
\subsection{Physics faculty views about different problem types}
\vspace*{-.35cm}

A prior study regarding physics instructors' views about different problem types in which they were presented with the same variations of a physics problem given to the TAs in the current study \cite{Edit}. It was found that the instructors generally valued different problem types intended to develop expert-like problem-solving but they were not as likely to use certain problem types. In particular, instructors' views about multiple-choice problems for introductory physics were not typically positive--the majority of faculty reported that they would never or rarely use multiple-choice problems, and the only reported use was for high stakes exams \cite{Edit}. Many faculty reported their reluctance to use multiple-choice problems was because it hindered their ability to monitor their students' thinking and they could not see their students' work. This finding regarding multiple-choice questions agrees with other research indicating that many faculty members never used multiple-choice questions, even for formative assessment purposes \cite{Dancy}. Regarding context-rich problems, it has been found that physics instructors generally valued context-rich problems and felt that such problems supported the goal of developing students' ability to plan and explore solution paths. However, they were not very likely to use context-rich problems to avoid stressful situations for students since they are complex and ill-structured and that such problems lacked clarity \cite{Edit}.

\vspace*{-.35cm}

\subsection{Prior work on TAs' professional development and their views about teaching and learning}
\vspace*{-.35cm}

Several studies have investigated TAs' views about teaching and learning \cite{Grading, Henderson, Rubric, Lin}. Prior research suggests that there are discrepancies between physics graduate TAs' perceptions of what teaching strategies are beneficial for students' learning and many of the findings of physics education research \cite{Grading, Henderson, Rubric, Lin, Marshman,Marshman1}.  For example, TAs have been found to struggle with the idea that effective grading practices can be a formative assessment tool, e.g., grading practices that encourage students to show their work can improve their learning from problem-solving and encourage them to learn from their mistakes \cite{Grading,Henderson,Rubric, Marshman}. Another study involving TAs' beliefs about example solutions provided to students shows that many TAs were unlikely to identify features in the problem solutions that the research literature describes as supporting learning goals for students \cite{Lin}. 

It has also been found that TAs' beliefs affect their teaching practices \cite{Goertzen, Chini}. Because of their role in decision-making on use of various problem types, both in the TAs' current capacity and in possible future roles as faculty, their beliefs about the pros and cons of posing an introductory physics problem in different ways and in different instructional contexts can affect the ways in which they use various types of problems to achieve different instructional goals. Thus, identifying the views of the TAs about the way in which a problem is posed can be useful in developing activities to improve their professional development and help them recognize the pedagogical value of posing the same problem in various ways to meet different instructional goals. 

\vspace*{-.35cm}
\subsection{Multiple-choice problems as a formative assessment tool}
\vspace*{-.35cm}
Multiple-choice questions may be used in both summative and formative assessments. Assessments that measure the extent to which students have learned and the goals of a course have been achieved at the end of a course, and nothing more, are called ``summative.'' 
On the other hand, ``formative'' assessment is an assessment in which both instructors and students receive feedback on students' understanding and their skills at a given point in time and there is opportunity to address student difficulties and help them learn those concepts better and improve their problem solving, reasoning and meta-cognitive skills. Formative assessments are often ``low-stakes;'' they are used frequently to actively engage students in the learning process, but have little impact on a student's final course grade. When multiple-choice problems are used as a formative assessment tool, they are often implemented in class as a low-stakes assessment and include ``distractor'' options among the given answer options.  ``Distractors'' are choices that are meant to be selected by someone who is not knowledgeable about the correct answer and for good multiple-choice questions focus on the common student difficulties found via research \cite{mashood,Engelhardt, lin,singh,antti,energy,gauss,rot,qmstest,Fuhrman}.  The presence of distractor choices reduces the chance that students can narrow down the correct answer based upon test-taking strategies rather than based on sound understanding of the content and problem-solving process \cite{Engelhardt}. If these distractor choices are based upon common student difficulties, the multiple-choice assessment can serve as a diagnostic tool to measure student understanding at a given point of time so that the instructor can address those difficulties using suitable pedagogical approaches \cite{Treagust}.   

There are many ways in which multiple-choice questions can be used in class to actively engage students and provide formative assessment feedback. For example, the use of multiple-choice questions in the form of clicker questions has been shown to enhance students' conceptual and quantitative problem-solving and reasoning skills \cite{Mazur} even in very large classes. When clicker questions are combined with active discussion, a majority of students have been found to have a positive attitude about the usefulness and enjoyment of the clicker questions and recognize the role of those questions in supporting their learning, in addition to being able to co-construct knowledge with peers \cite{Keller, Perkins, Perkins1,Sayer2}. If carefully sequenced, clicker questions have been shown to yield significantly higher conceptual understanding and help students feel actively involved in the learning process \cite{Lee}. In using clicker questions during in-class formative assessment, instructors can award partial or full credit to students simply for their participation in the clicker question activity (and not on whether the answers are correct).  Such a way of handling clicker questions has been shown to enhance discussion among students and more accurately assess student understanding \cite{James, Willoughby}.  

Peer Instruction is an example of a method which often utilizes conceptual multiple-choice questions as formative assessment tools and may positively impact students' conceptual understanding and problem-solving skills \cite{Mazur}.  However, clicker questions may be conceptual or quantitative.  Another application of multiple-choice questions in formative assessment is in the context of the ``inverted'' or ``flipped'' classroom, in which there are opportunities to incorporate in-class group problem-solving into lectures because the content which is normally covered in lectures to assigned videos and/or tutorials outside of the class \cite{Zappe,Karim,Karim1}.  Such in-class group problem-solving could include individual quantitative clicker questions which are designed to reinforce the content learned outside of class and provide efficient low-stakes formative assessment of students' mastery of material \cite{Zappe}. 

Low-stakes use of clicker questions has been found to reduce the achievement gap between underrepresented students or students with lower levels of prior knowledge and the majority students \cite{Lorenzo, Bonham, beichner}. This is because while all students benefit, those from an underrepresented group and/or with lower level of prior preparation benefit disproportionately more compared with other students \cite{Lorenzo, Bonham, beichner}. Thus, even in large enrollment classes, carefully-designed multiple-choice questions can be convenient for gathering efficient feedback regarding students' current knowledge and difficulties and can facilitate effective formative assessment to improve learning.   

Furthermore, interactive learning experiences can involve multiple-choice questions to keep all students actively engaged in the learning process. For example, in using interactive lecture demonstrations (ILDs), students are asked to predict what will happen before a demonstration takes place in the form of a multiple-choice question \cite{Sokoloff}.  Quick feedback on students' predictions for such ILDs can be gathered, even in very large classes, via clickers, flashcards, or even a show of hands if they are asked for such predictions in the form of a multiple-choice question \cite{Manivannan}. The ILDs have been associated with conceptual learning gains \cite{Sokoloff, Manivannan, Sharma}.
Multiple-choice questions may also be used as part of self-paced learning tools, e.g.,  self-paced learning tutorials, to give instant feedback to students based upon difficulties, and the results could also be followed up via in-class or online instruction that takes into account the students' difficulties found through their responses to the questions \cite{Putt, Keebaugh,quilt2, gausstut,Saul, Singh4,Singh41,Singh42}.  When students engage with such an approach to self-paced tutorials, student understanding is enhanced \cite{DeVore,DeVore1,DeVore2}. Similarly, well-designed multiple-choice questions that incorporate common student difficulties can help facilitate a ``Just-in-Time-Teaching'' (JiTT) approach wherein pre-lecture feedback is gathered just before instruction takes place and serves to guide the instructor in addressing student difficulties. Such ways of using multiple-choice questions as a low-stakes formative assessment tool have been found to enhance student learning 
\cite{Sayer}.

\vspace*{-.35cm}
\subsection{Context-rich problems as tools for promoting effective problem-solving skills}
  \vspace*{-.35cm}
  
Context-rich problems have been shown to be effective in helping introductory physics students become good problem solvers \cite{Heller}. Physics problems which are context-rich are often complex and lacking in structure, frequently provide redundant information or are missing information and have real-life contexts \cite{Heller}. One prior investigation suggests that students who worked in groups were more likely to use effective problem-solving strategies and show positive inter-dependence when working on context-rich problems than when working on analogous traditional textbook problems \cite{Heller}. Research also suggests that students who engage with context-rich problems are more likely to think about the concepts first, use diagrams in their problem solving process and have a more positive attitude about problem solving \cite{Ogilvie,Ogilvie1}. As students become more experienced in solving context-rich problems, they show progress towards expert-like problem solving \cite{Antonenko}. Because context-rich problems require students to formulate the question and make inferences, a systematic, expert-like problem-solving approach is more effective \cite{Reif}.  Thus, context-rich problems can facilitate progression towards expertlike problem-solving \cite{Reif} such as executing a conceptual analysis and planning of the problem solution before implementing the solution. Reflection and metacognition, as well as utilizing a well-organized knowledge structure, also play key roles in solving the problem and learning from the problem-solving process. \cite{Reif, Chi, ding,beichner,Schoenfeld}.

\vspace*{-.55cm}
\section{Methodology}
\vspace*{-.35cm}

{\bf Participants and description of TA professional development program:} A total of 97 TAs from a typical large research university participated in this study during 4 different years. Participants were physics graduate students who had teaching responsibilities (introductory recitation or lab instruction, and a majority were also assigned to help students in a physics tutoring center) and were concurrently enrolled in a mandatory TA professional development course that met once per week for 2 hours for an entire semester.  The TAs were expected to do approximately one hour of homework each week pertaining to the professional development course, in which various activities took place throughout the semester. During the course, initial activities related to the course focused on some general issues related to physics teaching and learning, e.g., discussion of some physics education research papers on common student difficulties in introductory physics. The discussion of grading practices occurred near the beginning of the semester, followed by discussions of pedagogy, including the use of tutorials and clicker questions as learning tools and the importance of integrating conceptual and quantitative learning.  After this, discussions turned to how different problem types (e.g., multiple-choice problems, context rich problems, problems that are broken into sub-problems, and traditional textbook style problems) can help students learn physics and can be useful in different instructional setttings to meet different instructional goals. Before the activities, TAs were involved in evaluating the effectiveness of multiple-choice questions on several standardized conceptual physics surveys, and predicting which choices students might select and why.  This activity also gave TAs the opportunity to reflect on the design of conceptual multiple-choice questions, and anticipate challenges their students might encounter.
The TAs also were given a physics problem and asked to present the solution to the TA professional development class as they would in their recitations. These presentations were video-recorded so that they could reflect on their teaching and also receive feedback from other TAs and the instructor.  Thus the problem-type activity was one of a number of activities all aimed at improving the professional development of the TAs and focused on investigating their views about physics teaching and learning.

\begin{figure}
\begin{tabular}{c}
\includegraphics[width=0.6\textwidth,angle=-90]{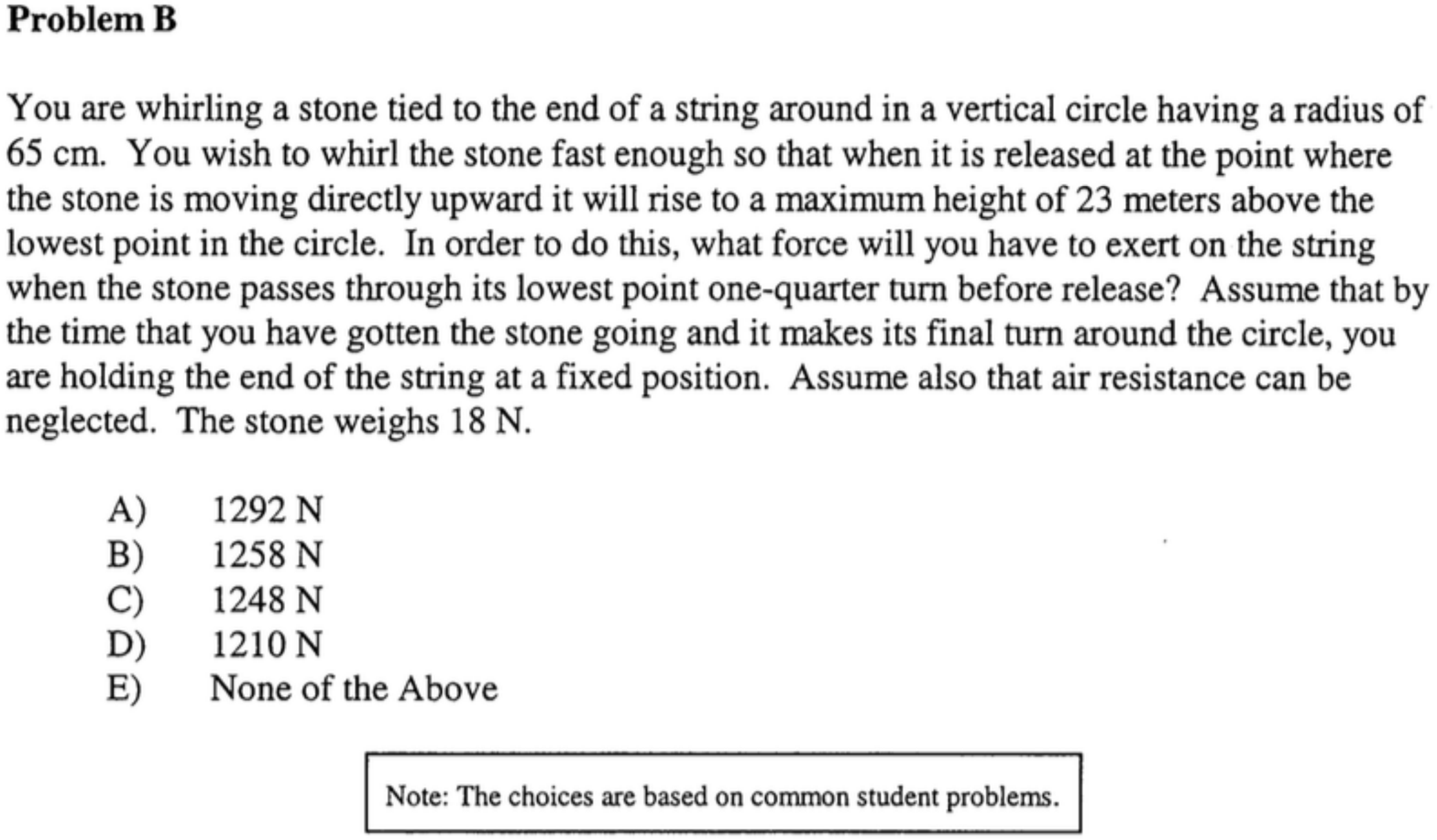}
\\
\includegraphics[width=0.6\textwidth,angle=-90]{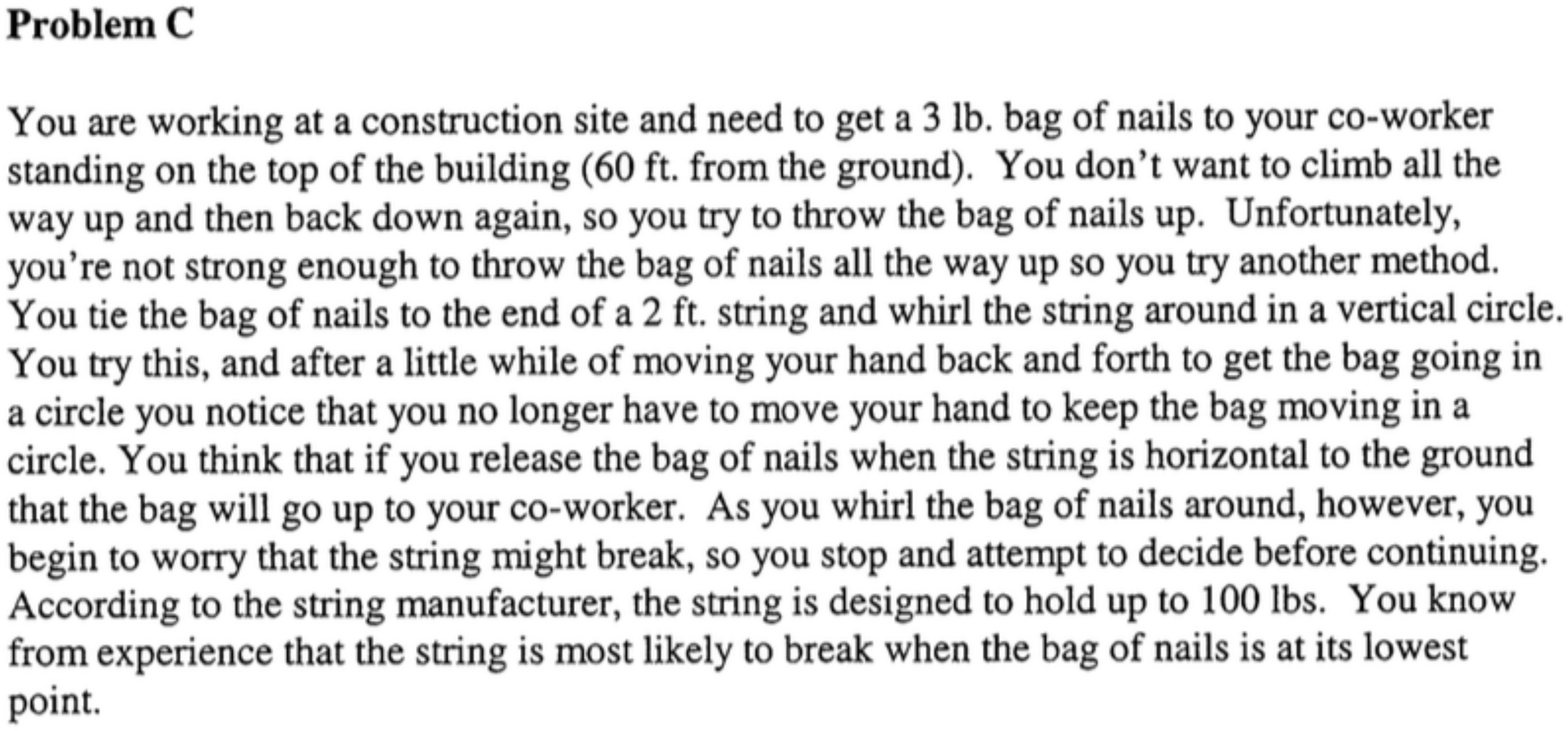}
\end{tabular}
\caption{\small The multiple-choice and context-rich example problems given to the TAs to illustrate these problem types}

\label{fig:problemvariations}
\end{figure}

{\bf Data collection tools and artifacts:} 
The data collection tools consisted of instructions and five example introductory physics problem types that had been developed previously to illustrate each problem type \cite{Edit}. The example problem types were designed for an introductory physics problem scenario in mechanics and served as a guiding example to illustrate what a particular problem type could look like for a given scenario for the activities. 
They included two different versions of a problem which was broken into sub-parts (one was framed in a more conceptual manner than the other), a multiple-choice problem, a context-rich problem and a traditional textbook version of the problem. Here we focus on the multiple-choice and context-rich problem types for which the example problems given for reference are shown in Figure \ref{fig:problemvariations} (Problems B and C, respectively). We note that it was made clear to the TAs several times that these were merely single concrete examples of a multiple-choice or context-rich problem type for illustration purposes and that they should, in general, reflect upon the instructional benefits of well-designed multiple-choice or context-rich problems.
For the example multiple-choice problem posed in a standard ``textbook'' style with choices that include common student difficulties as strong distractors,  there is a note at the end of the problem that the TAs could see and which was pointed out in discussion, explicitly stating that the choices are based on common student difficulties. 
We also note that for the multiple-choice problem type, although TAs were given a quantitative problem as an example, they were explicitly told to consider instructional benefits and usage of well-designed conceptual or quantitative multiple-choice problems in different instructional settings to meet different instructional goals. Thus, while the multiple-choice problem type example presented to TAs was a quantitative problem, the discussion in the TA professional development class focused on the merits of well-written multiple-choice problems in general, including the use of qualitative multiple-choice questions geared towards probing conceptual understanding. Similarly, even though a single example was given, the broader discussion centered on the merits of a well-designed context-rich problems.
As  illustrated in the example context-rich problem, this problem type typically requires that students first construct a concrete question and then solve the problem using the relevant information provided (extraneous information is also included and the problem requires explicit calculation after formulating it, as is typical of this type of problem).

Based upon our research questions, the TAs were asked to answer questions about these problem types on a worksheet in which they were
directed to list pros and cons for each problem type.  Specifically, in the instructions, TAs were asked to list at least one pro and one con for each problem type based upon the features each of the five problem examples contained. Data were collected over four different years. In the most recent year's worksheet, TAs were also asked what they would change about the example problems. In addition they were asked to rank the features of problem types on their instructional benefit (i.e., how instructionally beneficial the TAs judged each problem type to be), and to rank the problem types in terms of the level of challenge (i.e., how difficult the TAs judged each problem type to be for students). They were also asked how much they liked the problem types, and the likelihood that they would use the problem if they had complete control of the choice of the problem types to use. 
For example, a TA who ranked a problem 1 for ``challenging'' judged this problem to be the least challenging for students; a 5 for ``challenging'' indicates that the TA perceived it to be the most challenging among the five problem types. The rankings allowed us to investigate research questions 1 and 2. Furthermore, throughout all four years, TAs were asked to list pros and cons of the problem types. These pros and cons were useful for investigating why TAs ranked the problem types the way they did (research question 3).

{\bf Data collection in the TA professional development course and later in individual interviews:} TAs were given the example problem types and worksheets  
in the professional development course in the middle of the semester, when they had some teaching experience, in order to elicit their ideas about different problem types. 
They were asked to answer worksheet questions under the assumption that they had complete control over the introductory physics class, including control over problem types chosen for various purposes. 
The worksheet 
was completed as part of a homework assignment. 
Later, 12 participants who had taken the TA professional development course earlier volunteered to be interviewed in a one-on-one setting using a think-aloud protocol. These interviews took place at least one semester after the initial activity described here in the TA professional development course and were audio-recorded. Since TAs who answered the written questions may have become focused on example problems given to illustrate each problem type, TAs who participated in the interviews were asked questions both about the example problem type and about the problem types in general similar to the broader in-class discussion about the instructional benefits and pros and cons of well-designed multiple-choice and context-rich problems. Thus, these interviews served to more deeply probe the TAs' reasoning behind their written responses and to explore such questions as the use of well-designed multiple-choice (both conceptual and quantitative) and context-rich problem types in general.

{\bf Coding TA responses:}  Two of the researchers met weekly to identify appropriate coding categories for pros/cons; agreements on these were reached through discussion.  The researchers used open coding of the data from the individual homework assigned in the middle of the semester regarding the TAs' views of the problem types. The categories coded were arrived at over several weeks based on emergent themes.  Some categories were merged if they were found to be sufficiently similar. The inter-rater reliability was examined for the coding of the pros/cons in the year 3 data set, and the average Cohen's kappa \cite{Cohen} was calculated to be $\kappa = 0.982$. The most common pros and cons for the multiple-choice and context-rich problem types, along with their definitions and  examples from TA worksheets, are included in Tables   \ref{table:codingmc} and \ref{table:coding}.  We note that for the context-rich problem type, some TAs' responses could not always be appropriately coded as a particular pro or con because they were negative but non-specific in nature.  For example, one TA response was, ``never assign this.'' We were unable to code this response, and others of this nature, into a specific pro/con category.

\vspace{-0.2in}

\section{Results}
\vspace{-0.1in}
\subsection{Rankings }
\begin{figure}[htb]
\includegraphics[width=0.55\textwidth]{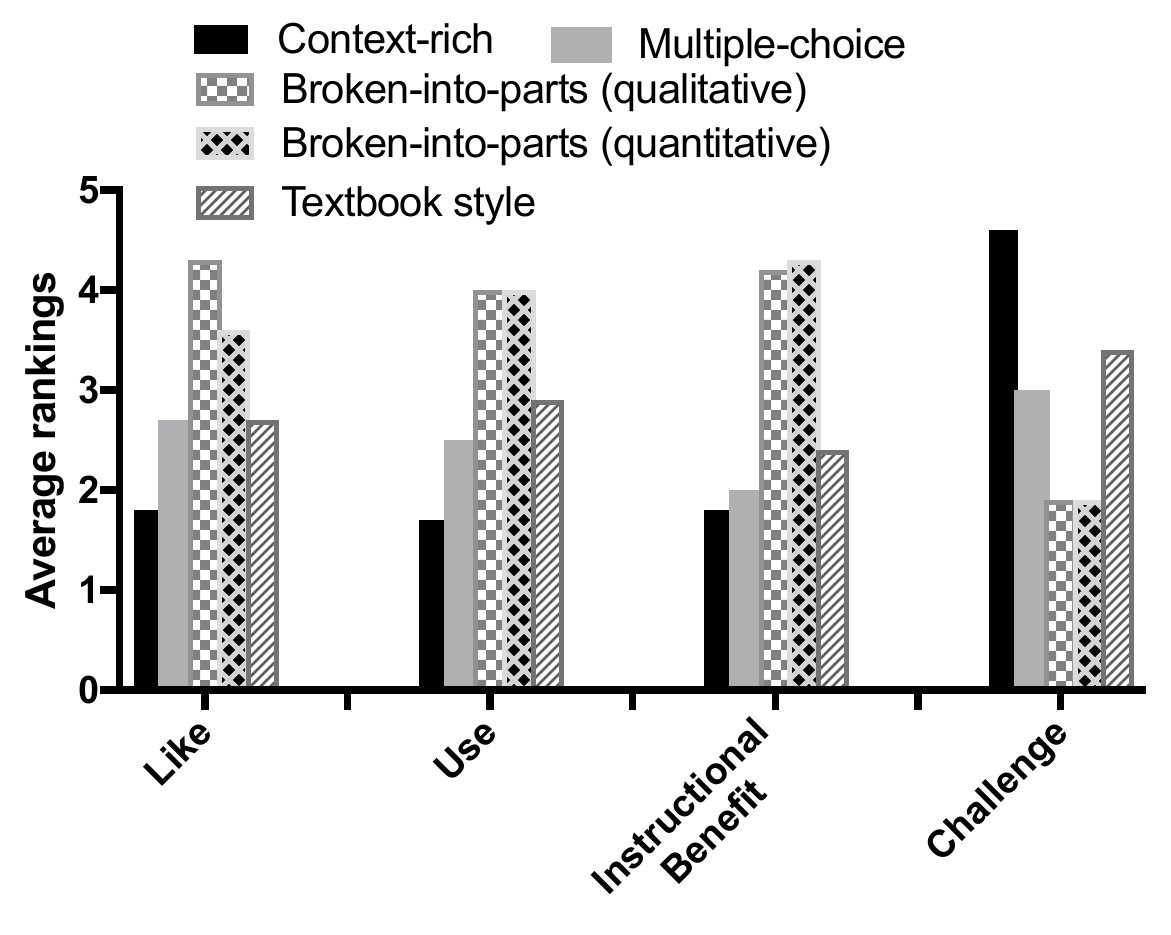}
\caption{\small Average rankings of the problem types presented to TAs}
\label{fig:ratingsall}
\end{figure}

\vspace{-0.1in}
We find that multiple-choice and context-rich problem types ranked low for like, use, and instructional benefit.
Figure \ref{fig:ratingsall} summarizes the ranking for all problem types in all four categories in order to compare the TA views of multiple-choice and context-rich problem types. In the category of ``instructional benefit'', TAs were asked to rank the \textit{features} of each problem type. In response to research question (1), the average rankings for ``instructional benefit,'' indicate that the multiple-choice and context-rich problem types are seen by TAs as the least instructionally beneficial of all example problem types.  The average rankings for ``challenge'' reveal that the context-rich problem type is ranked as the most challenging by TAs and that multiple-choice problem type is ranked as moderately challenging. The rankings for ``challenge'' are best understood in the context of the other problem types that the TAs were asked to rank. Two types of broken-into-parts problems were consistently ranked as the easiest two problems, and the context-rich problem was ranked as the most challenging. The multiple-choice and  standard ``textbook'' style problem types were ranked in-between these two extremes. In response to research question (2), the average rankings for ``use'' indicate that the multiple-choice and context-rich problem types were ranked the lowest of all problem types.  
Moreover, TA responses regarding the manner in which they would use multiple-choice or context-rich problem types indicate limited ways in which TAs would use such problem types.  
In response to research question (3), the pros and cons that the TAs listed, as well as their explanations of how they might use different problem types shed light on the rankings for these problem types.  

It is important to note that while written responses may have been guided by the example problem types given to the TAs to illustrate different problem types presented to TAs, TAs were explicitly asked to think not only about the specific example of a multiple-choice problem that they were given to illustrate a problem type, but also to think about well-designed multiple-choice problems in general including both conceptual and quantitative multiple-choice questions.  Moreover interviewed TAs responded in a similar manner to those who gave written responses for all of the preceding categories (i.e., ``instructional benefit,'' ``challenge,'' ``like,'' and ``use'') even when explicitly reminded to think more generally about well-designed multiple-choice problems including both conceptual and quantitative multiple-choice problems.  In other words, rather than the above rankings holding true only for the specific multiple-choice problem given, interviewed TAs saw little instructional benefit and were unlikely to use multiple-choice problems of any kind.  Thus, the interviews confirm that the rankings generalize to multiple-choice problems in general.

\vspace{-0.1in}
\subsection{Use of problem types}
\vspace{-0.1in}
As seen in Figure $\ref{fig:ratingsall}$, the TAs ranked the multiple-choice and context-rich problem types the lowest for ``use.''  
Reasons behind the use rankings can be found by examining their written responses on the worksheet regarding how they would use these problem types. Individual interviews and discussions also shed light on the fact that TA views about multiple-choice and context-rich problems were not restricted to one problem scenario given as an example problem to illustrate a problem type but were more general views.

While some TAs stated that they would never use a multiple-choice problems at all, some reported limited uses.  Of those who would use multiple-choice problems, the majority reported that they would use them only for quizzes or exams.  Interviews and written responses suggest that the TAs were very reluctant to use multiple-choice problems and, even if they would use them, they had a summative assessment view of such problems in mind.  
One TA explained his reluctance to use multiple-choice problems in an interview as follows: 
\textit{``I think multiple-choice takes the focus away from the problem at hand...This introduces more anxiety and confusion...I remember [one multiple-choice test] and I didn't do well, because I was so over thinking all of my answers and changing them multiple times.''}  This TA identified students' anxiety over getting the correct answer and evaluating the validity of the choices given as a reason for avoiding using multiple-choice problems.  It was clear from the discussion during the interview that this TA was thinking about a high-stakes, summative use of multiple-choice problems, and the impression that those situations left on him as a student for being anxiety-producing.  
Similarly, when other TAs also spoke about using multiple-choice problems, their responses focused on a summative assessment type of usage.  Another TA, when explicitly asked how he might use multiple-choice problems, said, \textit{``Maybe if I want to make the final totally multiple-choice,''} but did not offer any other type of use for such problems.  
Another TA offered, \textit{``Quiz or exam type things, not homework... I don't think it's necessarily helpful in a homework to have the multiple choices.''} These TAs (and a majority of others) also did not offer any other way of using multiple-choice problems, even when asked explicitly.  Most TAs, when asked to come up with at least one pro, noted that the main pro of well-designed multiple-choice problems was to help them with the task of grading the quiz or exam efficiently.

Indeed, few TAs  mentioned using multiple choice problems as a formative assessment tool such as ``clicker questions,'' in-class group problem-solving in flipped classes, or self-paced tutorials which can be used even in very large classes (all TAs were aware of clicker questions and other technology since it was discussed in an earlier session in the TA professional development class).  It is also important to note that TAs were specifically asked to assume they had full control over teaching the course independently when responding on their worksheets regarding problem types and their instructional benefits, etc. (and this point was also emphasized several times). 
Thus, the absence of possible formative uses in their written responses as well as in interview data suggest that the TAs did not immediately recognize these possibilities as useful for overall course instruction, not due to their limited roles as TAs. 
A lack of valuing of the multiple-choice clicker questions was observed in the comments made by TAs during interviews. For example, when the interviewer specifically asked about use of a multiple-choice question with clickers, one TA said that he does not like to think about using multiple-choice questions even for clicker questions.
Discussions with this TA suggest that he was well aware of the fact that clickers can be utilized in physics classes, but he did not appear to have reflected on the instructional value of using multiple-choice clicker questions as a formative assessment tool.

Regarding the context-rich problem, the ranking for ``use'' was the lowest of all problem types, indicating a common reluctance among TAs to use a context-rich problem. The majority of TAs did not envision any possible situation in which context-rich problems could be used effectively, e.g.,
 in a collaborative group problem solving session or as part of a homework assignment. Furthermore, even though the instructions regarding the worksheet completion included a directive to think of at least one way in which each problem type could be used, over 20\% of the TAs stated that they would never use context-rich problems in any way and listed no pro even when asked for at least one pro. 
Moreover, of the few TAs who mentioned group work as a potential use,
most did not mention that the group work could help actively engage students in co-constructing knowledge and learning problem-solving skills. 

Similarly during interviews, when asked how they might use the context-rich problems, many TAs noted that they would not use context-rich problems at all and only one of the TAs suggested the idea of group work as a possibility. Exhibiting this reluctance to use context-rich problems, one TA struggled to think of a possible way to use a context-rich problem, saying, \textit{``I think this might not be the kind of question you ask in a recitation...maybe this is the outline for a lab or something, but it doesn't seem right [even then].''}  Further discussion suggests that this TA felt that the context-rich problems were not well-suited to use in a recitation even for group work. Furthermore, even in a lab setting where time may not be as big a factor and where students might collaborate, the TA was reluctant to endorse using context-rich problems. Other TAs also had similar sentiments.  
Some possible pros and cons stated by the TAs for the multiple-choice and context-rich problems are listed in Tables \ref{table:codingmc} and \ref{table:coding}.

{\bf TAs' perceived cons of multiple-choice problems:}
One reason for TAs not using multiple-choice questions was a feeling that such questions constituted a ``trap.'' This theme of ``trapping'' students is evident in the following quote: \textit{``If I were to do a problem and get one of the answers... 
then you look at the others and you doubt yourself and you get started thinking about patterns in the question and what you're off by, or is it 'none of the above?' Is it all a trick?''}  This TA expressed that the alternative choices provided may throw a student off by the ``patterns'' that appear to them to exist in these alternative choices.   
Other TAs expressed similar sentiments.  The concern for perceived fairness is definitely laudable, but interviews suggest that it arises from TAs assuming a high-stakes assessment in which an incorrect answer may have a major impact on a student's grade. Interviews suggest that the assumption of high-stakes assessment often influenced the fact that the distractor choices were seen by some TAs as a ``trick'' rather than having instructional value in evaluating whether the student was truly knowledgeable about the correct concepts and problem-solving approach.

In addition, TAs did not view multiple-choice problems as reflecting student understanding.
Table \ref{table:codingmc} shows the most common pros and cons mentioned by TAs in written responses. As can be seen in Table \ref{table:codingmc}, a majority of TAs listed the con ``no parital credit/no understanding shown.''  This con referred to TA responses that indicated that they felt that this problem type would be a poor reflection of students' understanding, that the students may not get partial credit, and/or that the students could potentially guess the correct answer. Thus, this category encompasses several reasons why TAs might be concerned that the multiple-choice questions 
would not necessarily give students appropriate credit for their level of understanding.

For example, the issue of potentially guessing the correct answers for the multiple choice questions was mentioned by many TAs.  In an interview, one TA explained the concern about guessing as a reason for why this type of problem may not be valuable:\textit{``There's always a possibility that they could make a guess, and I don't really see any particular value to using multiple choice....''}  The possibility of guessing appeared to influence this TA's view of the problem's perceived value.  Guessing is certainly a valid concern and it is encouraging that TAs were concerned with the implications for fair grading when guessing is a possibility.  However, it appears that TAs did not think more deeply about how distractor choices or other elements of well-designed multiple choice questions might discourage guessing (e.g., consistent units and similar numbers might not allow a student to easily ``rule  out'' an answer choice).
It is interesting to note that sometimes TAs were reporting concerns that could be considered to be contradictory. In particular, TAs often disliked the idea of guessing, but also disliked the idea of using distractors which can dissuade guessing because they felt that the distractors constituted a ``trap.''  This suggests a possible lack of reflection and/or a lack of awareness of the instructional value of design elements such as distractor choices. It also appears that TAs did not think about how the formative assessment value of such questions could outweigh the risk of guessing for a low-stakes assessment use, such as clicker questions or as assessment in self-paced learning environments.

\begin{table*}
\caption{\small The most commonly listed pros/cons of the multiple-choice problem type and the percentages of TAs who listed them. Some TAs  listed more than one of the following or other pros/cons not listed here.}
\label{table:codingmc}
\centering
\fontsize{10}{10}\selectfont
\begin{tabular}{|m{8em}|m{8em}|m{15em}|m{5em}|}
\hline 
Code  & Definition & Examples & Percentage of TAs
\\
\hline

(Pro) check & students can check their answer & ``can check your answer'' & 22
\\
\hline
(Pro) time & saves time & ``quick way to get answer'' & 14
\\
\hline
(Pro) grade & easy/efficient to grade &``makes problem faster to grade'' & 9
\\
\hline
(Con) no partial credit/no understanding shown & can guess; does not demonstrate understanding & ``can't tell if they can do the process, can't prevent guessing'' & 54
\\
\hline
\end{tabular}
\end{table*}

\begin{table*}[!htbp]
\caption{\small The most commonly listed pros/cons of the context-rich problem type and the percentages of TAs who listed them. Some TAs  listed more than one of the following or other pros/cons not listed here. However, other pros were negligible and the overlap of the two cons listed below represents only 5\% of all TAs.}
\label{table:coding}
\centering
\fontsize{10}{10}\selectfont
\begin{tabular}{|m{8em}|m{8em}|m{15em}|m{5em}|}
\hline 
Code  & Definition & Examples & Percentage of TAs
\\
\hline

(Pro) real & real life scenario; can help students relate physics to their lives & ``gives insight to how physics laws and principles are ingrained in our lives'' & 47
\\
\hline

(Con) time & time-consuming & ``takes up a lot of time''  ``can take too long'' & 33
\\
\hline

(Con) unclear & problem is vague and/or confusing & ``can be ambiguous; not sure what to do''  ``hard to decipher'' & 42
\\
\hline
\end{tabular}
\end{table*}

In addition to guessing, some TAs mentioned the issue of partial credit. In an interview, one TA expressed this concern as follows: \textit{``If they did good work, and they chose b instead of a, and they got zero points, I wouldn't like that as a TA.  I expect myself to be someone who grades on problem solving merit, and the thought process as opposed to the final answer''}.  This TA was concerned about a student potentially getting a score of zero for a multiple-choice problem, which is commendable. Interviews suggest that such concerns were often based on an assumption that multiple-choice questions would be given in a high-stakes summative assessment.  The concern for fairness that this TA and others showed is laudable, but it appears that he did not realize that partial or full credit may be built into low-stakes formative assessments (as is often done, e.g., with clicker question responses). Other TAs expressed similar sentiments about the negative aspect of students not receiving partial credit. Many TAs assumed that multiple-choice problems necessarily mean that credit is not possible if the answer is not correct. However, credit possibilities do exist, such as participation or completion credit for clicker questions or in online assessments as part of a self-paced learning environment.

{\bf TAs' perceived cons of context-rich problems:}
Table \ref{table:coding} shows that the most commonly stated con was coded by the researchers to be in the category ``unclear.'' 
 ``Unclear'' refers to TA responses which described the problem feature as confusing (e.g.,``Overly confusing and frustrating'') or lacking in clarity or explicit question (e.g.,``The point of the problem is not clear'').  In fact, the con ``unclear'' appears to be one major reason why the TAs thought that the context-rich problem type was highly challenging and may have contributed to the TAs' reluctance to use this problem type or recognize its instructional benefit in any setting to achieve instructional goals. The majority of TAs who ranked the context-rich problem type as the most challenging and low in instructional benefit also mentioned a con categorized as ``unclear.''  Similarly, the majority of TAs who indicated they would be unlikely to use context-rich problem type by ranking it low for the category ``use,'' also mentioned a con coded as ``unclear.'' 

The TAs who listed the con ``unclear'' had a variety of reasons for why they perceived this type of problem to be lacking in clarity, and often the same TA stated more than one reason for why he/she felt this problem type is unclear. Written responses and interviews suggest that one reason the context-rich problem given as example was often viewed as unclear by the TAs was that there was a lack of an explicit question in it (which is common for problems posed in a context-rich manner).  For example, one interviewed TA explained that the lack of an explicit question made the problem confusing: \textit{``This one is very vague.  It's not asking any question, so the student might get confused about what they're supposed to do....''}  This type of sentiment about the lack of explicit question and the problem being ``vague'' or a source of confusion for students was commonly mentioned by the TAs both in written responses and interviews. Another interviewed TA who described the example problem as ``unclear'' explained, \textit{``at least if you give a student a problem where they know what they need to do, they can ask you questions about that, but I can imagine students sitting there saying `I don't know where to start.'''}   Moreover, discussions with this TA suggest that he felt that his students will not be able to proceed with any context-rich problems. He noted that without being able to discern a clear goal, the students will neither be able to make sense of the problems nor be able to reach out for support from the TA or instructor, as they will not know what questions to ask. Similar to this TA, many TAs connected a perceived lack of clarity to the fact that the context-rich example problem does not ask an explicit question.  Another TA stated, \textit{``It is challenging... because the wording of the problem is vague. I don't even really see an explicit question that it's asking.''} This TA (and many others) explicitly made a connection between the level of challenge and the lack of a concrete question posed. Discussions with TAs suggest that many felt that it did not make sense to give these types of problems to students under any circumstance they spontaneously came up with in the moment or that were mentioned by the interviewer. These TAs appeared not to recognize the instructional benefit of requiring students to determine the questions that are being asked in context-rich problems, which is the rationale for not including questions in context-rich problems.

Another reason many TAs found this problem type to be ``unclear'' was that the TAs felt it was verbose with too much extraneous descriptive information and that their students will have difficulty interpreting it (note that wordiness and possible extraneous information are common features of context-rich problems).  This reason for perceived lack of clarity also seems to be contributing to the high ranking for challenge and low ranking for instructional benefit and likelihood to use. For example, one interviewed TA first described the problem type as having ``too much detail,'' and then went on as follows, \textit{``it's very good that it has a story to it but the story is too much and the science is not enough...because this is a physics class and... it looks more like a story than a physics problem.''}  It is interesting that the TA appeared to view the narrative aspect of the problem type (i.e., the ``story'') as separate from the ``science'' of the problem. 
This TA appeared to be of the opinion that the context-rich problem type may not hold benefit for physics students, since he felt that this sort of problem type does not belong as part of a physics class. 
Further discussion with the TA suggests that he preferred textbook problems and viewed the detailed descriptive narrative in a physics problem as being outside of the scope of what should be part of a ``usual'' physics curriculum. This could explain why this TA reported that he would not be likely to use context-rich problem type. Another TA remarked, \textit{``I think... intuition in being given a word problem and knowing how to translate that into math is the hardest part.''} Further discussions with this TA suggest that he agreed that formulating the problem (i.e., translating the problem from words to quantitative expressions) may potentially be useful for students.
However, he then stated that he would never use this type of problem in his own class due to the fact that it is unclear and challenging for students and his students would not know how to solve these types of problems. Similarly, another TA who ranked the context-rich problem type the lowest for likelihood to use such problems said: \textit{``It's not about the physics; it's just about the wording... I find this question itself is kind of testing how you understand a paragraph of description, but the physics itself might [not] be...''} This TA felt that the wording of the problem as a descriptive paragraph was something separate from the physics involved, almost as if implying that the paragraph form obfuscates the underlying physics.  Indeed this TA went on to say that \textit{``It's kind of confusing...I took a long time to understand it...It's not quite clear.''} This TA mentioned both the lack of an explicit question and the lengthy narrative as problematic for the problem's clarity, and further mentioned that he did not see instructional benefit in using context-rich problems in general.

In addition to the con ``unclear,'' another common con category was ``time''. ``Time'' was the category used for TA responses that indicated that solving this problem type would be too time consuming for introductory students (``It may take a while for the student to interpret the problem into a mechanics problem'') and/or did not make the best use of time (``Takes time reading things that are not directly helpful for solving the problem''). The con ``time" may also have contributed to TAs' reluctance in using this problem type in their own classes. Of the TAs who ranked the context-rich problem type the lowest in terms of ``use", the majority listed ``time'' as a con. A TA's reluctance to use a context-rich problem due in part to time constraints is evident in the following comment during the interview: \textit{``The student will take much more time to solve this...  I think the student has to read this question so many times more to understand it.  ...I won't use this at all.''} We note that this TA identified both the time it will take the student to solve the problem and also the time required to read and understand the problem statement as being problematic with regard to the use of this problem type. Further discussions suggest that, in general, he would not use this problem type in his classes. In a similar manner, another TA who reported that he would never use a context-rich problem stated: \textit{``The students have to spend too much time trying to figure out what they are supposed to answer...It does not help the student at all.''}  This TA identified the time needed to interpret this type of problem and construct a well-defined question as excessive and went on to say that this time requirement makes the problem type unhelpful to the student. He further noted that the complexity negatively impacts the instructional benefit of this problem type in his mind.

\vspace{-0.15in}
\subsection{With regard to the overall instructional benefits, the pros do not appear to outweigh the cons}
\vspace{-0.1in}

 Many TAs did not list any pros for the context-rich problem type even though they were asked to list at least one pro for each problem type, and the pros listed for the multiple-choice problem type appear to be mainly centered around logistical issues.
 The inability to come up with a pro for the context-rich problem type and the logistical nature of the pros given for the multiple-choice problem type corroborates the rankings for ``instructional benefit,'' a category in which multiple-choice and context-rich problem types ranked the lowest. 
This suggests that TAs did not perceive these types of problems to be instructionally beneficial.  Moreover, the paucity of compelling pros for these problem types may explain the low average rankings these problem types received for ``like'' and ``use.'' 

\textbf{TAs mostly cited logistical issues as pros for using multiple-choice problems:}
Although none of the individual pros were mentioned by a majority of TAs (i.e., the most common pro was only mentioned by 22\% of TAs), the common pros included several sentiments that appeared to center around logistical issues. None of the pros appear to relate to instructional benefits such as the use of multiple-choice questions as a formative assessment tool. In fact, the most common pros had a theme of logistical considerations.  The perceived time-efficiency, the ease and quickness of grading, and the idea of students checking their answers can be viewed as logistical concerns more than perceived instructional benefit. This finding suggests that the positive aspects TAs saw in the multiple-choice problem type did not have to do with such problems benefiting student learning when used as a formative assessment tool. 
Furthermore, it should be noted that none of the pros listed in Table \ref{table:codingmc} were stated by even one-quarter of the TAs.  This is because almost 40\% of TAs did not list pros at all, even when explicitly asked to list at least one pro of well-designed multiple-choice questions in some instructional setting to achieve some instructional goals.

Some TAs expressed that ease and/or efficiency of grading multiple-choice problems was a pro (and most often the only pro).  The researchers coded these responses in the category ``grade.''  As an example of this sentiment,  
one TA mentioned the ease with which a multiple-choice problem can be graded as a positive aspect of the problem type. As mentioned earlier, the code ``grade'' included any time-savings on the part of the TA in grading the multiple-choice problem type. A TA response indicating this time-efficiency when it comes to grading was \textit{``Time saving to check.''}  Both the ease and efficiency of grading a multiple-choice problem type were commonly considered as pros by the TAs.  One TA, who mentioned in an interview that the multiple-choice problem type had the perk of being easy to grade, acknowledged this as follows: \textit{``It's not a nice answer but it's pragmatic.''} Logistical issues had obvious appeal to the TAs, but did not imply any deeper instructional benefits of multiple-choice problems as a formative assessment tool.

Another common pro listed by the TAs was coded in the category ``time.'' In this case, ``time'' means that the TAs felt that multiple-choice questions would be time-efficient or would save time for the students.  We note that the pro ``time'' specifically applies to only TA responses that indicated that the time-savings would be for the students (as noted earlier; responses that indicated that the multiple-choice problems would be faster or easier to grade, were coded as ``grade'').  For the pro ``time,'' TAs explained that not needing to show work in order to select a choice would make the problem-solving faster.  
Expressing the sentiment of time-savings, one TA succinctly noted that the multiple-choice problem type would be \textit{``[A] quick way to get answer.''}  Similarly, another TA stated that it \textit{``Saves time to get correct answer.''}.  This perceived time-efficiency was often cited as helpful for a quiz or exam situation where time might be limited.  One TA expressed this sentiment by stating: \textit{``Saves time for a quiz.''} 
However, if the multiple-choice problem was quantitative (as in the example to illustrate the problem type), 
the problem-solving process may still require a comparable amount of time as analogous problems posed in other ways. The TAs often missed this point.

Another common pro was ``check,'' which refers to TA responses that indicated that they viewed  multiple-choice problem type as feasible to check the correctness of one's answers.  For example, one TA explained that having the choices present means that \textit{``[Students] can see that their answer might be correct.''}  A similar idea was expressed by another TA who stated that: \textit{``If the student makes a calculation error, they will know right away.''} And yet another TA simply stated that the main benefit of the multiple-choice questions is that \textit{``You can check your answer.''}  TAs with such responses often regarded this feature as a positive aspect of the problem type--they felt that the students could benefit from checking if they had the correct answer.  However, TAs did not express or recognize that the presence of common incorrect answers included in the choices could preclude students from checking that an answer was correct, since they may simply be verifying an incorrect answer choice (one which is based on a known common student misunderstanding). It appears that TAs did not look past the obvious feature of one of the choices being correct to realize that carefully-designed distractor choices would make it difficult to check one's answer.

\textbf{Only one common pro listed for context-rich problem type, and it was not viewed as compelling:}
Of the pros TAs did mentioned for the context-rich problem type, the most common pro was ``real'' 
(i.e., relatable to a real-life scenario).  One TA put it succinctly: 
``Connects to daily life''.  We note that the TAs seldom mentioned any other pro. Interviews suggest that the pro ``real'' may not be perceived as compelling enough to outweigh the negative light in which the TAs viewed the context-rich problem type overall.  For example, one interviewed TA stated, \textit{``There's some redeeming elements to it, like I like that this frames it from the perspective of the student, so they can think about what they see and feel while they're whirling the string around.  So it's not all bad, but...I would tend not to use it at any level. I just don't think it's an effective problem.''}  This TA appeared to recognize some ``redeeming elements'' in the real-life aspects of the context-rich problem, but qualified this pro by stating that this is not enough of a reason for him to use a problem like this in his own classes in any way.  Additionally, the remark about it not being effective appears to speak to his low opinion of the problem in terms of its instructional benefit. Further discussions suggest that this TA would not use context-rich problems in general in his classes. Other TAs had similar views.

Thus, written responses and interviews suggest that  the TAs, in general, appear to have a negative opinion of the context-rich problem type. The total percentage of TAs who listed one or more significant cons to the context-rich problem is 80\% and other negative responses could not be coded since they were not specific.
Indeed, it is interesting to note that the context-rich problem type elicited a strong negative response from many TAs. Some TAs were extremely negative with statements such as: ``Absolutely does not help the students at all'', ``meaningless'', and ``It sucks''.

\vspace{-.15in}
\section{Discussion and Summary}
\vspace{-.1in}

Overall, TAs had major concerns about the instructional benefits and the use of multiple-choice and context-rich problems. They did not recognize the substantial possible instructional benefits of both of these problem types. Since TAs are not blank slates, their own prior experiences as students may have played an important role in their perceptions of the instructional benefits of different problem types. 
In response to research question (1) and (2), we find that, in general, most TAs viewed multiple-choice and context-rich problem types as challenging, but did not perceive them to be instructionally beneficial under any circumstance they spontaneously came up with in the moment or that were mentioned by the interviewer. TAs reported that they would not be likely to use them. Both in the written responses and in the interviews, most TAs mentioned using multiple-choice questions only for convenience in summative assessment, and most could not think of any instructional purpose for which they would use context-rich problems even though they were asked to list at least one instructional benefit in some instructional setting.  They did not mention multiple-choice questions as playing an important role in formative assessment, nor did they appear to envision using context-rich problems to engage students in collaborative group problem-solving. 

In response to research question (3), we found several reasons for why TAs viewed these problem types the way they did. Concerns TAs mentioned were very legitimate, e.g., one does need to think about the possibility of a student guessing on multiple-choice questions or one must find time to allow students to solve a context-rich problem that requires interpreting and formulating the goal of the narrative first.  
However, most TAs did not readily identify many of the instructional benefits of either the context-rich or the multiple-choice problem type in different instructional settings to meet various instructional goals. The TAs were specifically told to assume they had complete control of a hypothetical introductory physics class.  Also, even if in the written responses, they focused heavily on the example problem types given for illustration (although they were asked to consider the problem types in general),  since class discussion and interview data in which TAs were specifically asked to think about these problem types in general corroborates these findings, lack of recognition of instructional benefits of these problem types appears to be due to TAs existing model for teaching. Since TAs are not blank slates, their views about teaching and learning may be heavily influenced by their own experiences as students. These findings  agree with findings from previous studies regarding TAs' beliefs about assessment and example problem solutions, wherein TAs struggled to identify aspects that were supported by the research literature as beneficial for students \cite{Grading, Henderson, Rubric}. 

Indeed, the use of multiple-choice problems as a low-stakes formative assessment tool appeared not to be recognized by the TAs, where this type of use lends itself to efficiently identifying the common difficulties students have 
and provides  an opportunity for students to reflect on their own difficulties. 
  It is telling, for example, that a significant number of TAs reported that they would never use multiple-choice questions in any way even if they had access to well-designed multiple-choice problems, and that, among those who would use it, they primarily had summative assessment and logistical issues in mind.  The idea of utilizing the information one could gather from multiple-choice questions, e.g., using clickers in a large classes to figure out the fraction of students who had a common difficulty,
in order to use appropriate curricula and pedagogies to address those difficulties appeared not to occur to the vast majority of TAs. In other words, TAs did not mention any of the plethora of ways that multiple-choice questions can be used in low-stakes formative assessment, even in very large classes. The use of multiple-choice questions as formative assessment tools is not only restricted to clicker questions, which are very versatile for both individual or group questions and both conceptual or quantitative problems, but such questions can also be integrated with interactive lecture demonstrations, self-paced tutorials,  pre-lecture videos and corresponding assessment tools and with other ``Just-in-Time-Teaching'' strategies. Even when explicitly asked for at least one pro for the multiple-choice questions, the power of such questions as a formative assessment tool was not immediately recognized by a majority of the TAs. On the other hand, it has been shown that multiple-choice questions can be implemented as effective formative assessment tools and implementation of well-designed multiple-choice questions can make them excellent low stakes assessment tools when well-designed questions with good distractor choices based upon research on student difficulties are used, and  students are awarded participation credit regardless of the correctness of their answers, etc.)  \cite{Mazur, Putt, Keebaugh, quilt2,gausstut,Saul, Bonham, beichner, Lin3,Lin31}. For example, the use of good distractor choices or sequences of multiple-choice questions which build on each other can help reinforce concepts, encourage a desirable problem-solving approach, and/or dissuade guessing.  Yet, to many TAs who discussed the presence of the distractor choices in detail during interviews, the instructional value of these distractors did not appear attractive in any instructional setting they could think of to meet instructional goals in a low stakes context.  In fact, some TAs expressed serious misgivings that these served to simply maliciously ``trap'' a student. Concern for fairness is certainly a very positive aspect of instruction, but it appears that the TAs had an underlying assumption of a high-stakes assessment use of multiple-choice problems and a lack of awareness of low-stakes formative assessment and the instructional value of distractor choices (e.g., you can award full credit for clicker questions for participation only regardless of the correctness of the answers).  

Similarly, TAs did not mention the use of context-rich problems for collaborative group problem-solving. Moreover, many of the perceived drawbacks to using context-rich problems mentioned by TAs are identified as positive aspects of context-rich problems in the  research literature \cite{Heller, Ogilvie}. Context-rich problems are usually purposely designed with these features because of the benefit that those features afford for helping students actively engage in learning effective problem solving strategies, including the importance of doing a qualitative analysis and planning of the problem before jumping into the implementation phase.  For example, the lengthy narrative without a clear question can develop students' ability to differentiate what information is important, identify the relevant concepts, and formulate the problem.  Similarly, the length of time needed to solve a context rich problem is partly due to the fact that it is realistic. Not every problem one encounters in real life can be solved quickly, and the time required to solve a context-rich problem can also help students learn the importance of perseverance in problem-solving involving realistic situations. This is a lesson that could be beneficial for students, since research has shown that students often are likely to give up if they cannot solve a problem in 10 minutes \cite{Schoenfeld, Mason}. Yet these beneficial aspects were seen by a majority of the TAs in a negative light. 

It is worth noting that TAs were more positive and able to perceive instructional benefits of other problem types, as seen in Figure \ref{fig:ratingsall}. 
However, written responses, class discussions and interviews suggest that their perceived instructional benefits in other problem types (e.g., broken into parts problems) does not represent an understanding of the benefits of formative assessment.  Moreover, when it came to the multiple-choice and context-rich problem types, it was difficult for TAs to perceive their instructional benefits in any instructional setting even when asked for at least one such case.

We note that one limitation of our findings is that even though it was explained to the TAs that they should reflect on the instructional benefits and use of well-designed multiple-choice and context-rich problem types in general, they were given one example of each problem type to illustrate the problem type.  Thus, it is possible that the written data were heavily guided by specific examples given to illustrate each problem type even though the TAs were asked to consider the instructional benefits of problem types in general.  However, discussion in the TA professional development class and interview data provided opportunity to elicit perceptions of these problem types in general for well-designed problems to meet various instructional goals. Furthermore, the qualitative interview data provided reasons for responses for each problem type.  Also, while the multiple-choice example presented to TAs was a quantitative problem, the discussion in the TA professional development class focused on the merits of well-written multiple-choice problems in general, including the use of conceptual multiple-choice questions geared towards probing conceptual understanding. Thus, while one limitation to our findings is the ability to generalize the results since a single example illustrating each type of problem was given, the agreement with the interview data suggests that the written results are likely to be true for TA perceptions, in general. 
Moreover, the most recent year's worksheet included ``instructional benefit'' rankings that instructed TAs to rank the \textit{features} of the problem types first before ranking the problem types themselves, and the feature rankings by the TAs agrees with the TA rankings of the problem types themselves.
In other words, it was the \textit{features} of these problem types (e.g., multiple-choice, context-rich, broken-into-parts, etc.) that appeared to rank low for instructional benefit in the opinion of the TAs consistent with individual interviews and TA professional development class discussions on these issues.

 Our results are partly consistent with previous findings on faculty views which utilized the same example problems \cite{Edit}. In the faculty study, it was found that many faculty did not value multiple-choice problems and that either they would not use multiple-choice problems at all or they would only use them in testing situations for convenience in high-stakes summative assessment \cite{Edit, Dancy}.  Like the faculty in the previous study, we find that TAs assumed that multiple-choice questions offer only logistical benefits for time-saving in high-stakes summative assessment, and that the inability to see students' work was seen as a major reason to avoid using these types of problems.  With regard to the perception of the context-rich problem type, our results for TAs' views differ somewhat from instructors', who were found to value the context-rich problem type for its capacity to develop students' abilities to plan and explore solution paths \cite{Edit}.  Nevertheless, while faculty appeared to discern the benefits of a context-rich problem type more readily than the TAs in our study, the TAs do share some similarities of perception with those of faculty.  Specifically, like the faculty, the TAs in our investigation were not likely to use context-rich problems in their classes, both mentioning issues of clarity as problematic \cite{Edit}.  Faculty also mentioned avoiding student stress as a reason to avoid using context-rich problems \cite{Edit}, which was also mentioned by some TAs, but in smaller numbers than the issues of time and clarity. This suggests that while faculty may value a context-rich problem type more than TAs (or at least expressed in one-on-one interviews that such problems could be valuable for introductory students), a majority of faculty and TAs noted that they will likely not utilize context-rich problems in any instructional setting for instructional purposes.

Another limitation of this study is that  the TAs who participated in this study were from a typical large research university in the United States, so  our findings may not apply to other institutions which are different.  However, leaders of TA professional development programs at similar institutions can benefit from the findings of this study and develop activities to help TAs reflect on the benefits of multiple-choice problems as a formative assessment tool and of context-rich problems as a tool for actively-engaging students in collaborative group problem-solving to aid students in the development of problem-solving skills. For example, professional development programs can help TAs reflect on the plethora of potential formative assessment uses of multiple-choice questions, e.g., as clicker questions, as low stakes assessment in self-paced learning environment or in ``Just-in-Time-Teaching'' context, etc.  In addition, professional development programs can help TAs reflect on research that supports context-rich problems as instructionally beneficial for students (especially when used in collaborative group problem solving) in order to help them come to discern their instructional benefits to achieve various instructional goals.

\begin{acknowledgments}
We thank the National Science Foundation for awards DUE-1524575 and PHY-1505460. 
We are very grateful to all the students who participated in the study.
\end{acknowledgments}

\end{document}